\documentclass[twocolumn,showpacs,preprintnumbers,amsmath,amssymb,aps]{revtex4}
\usepackage{graphicx}
\usepackage{dcolumn}
\usepackage{bm}
\def\be{\begin{equation}}
\def\ee{\end{equation}}
\def\bea{\begin{eqnarray}}
\def\eea{\end{eqnarray}}
\begin{document}
\title{Rotating Scalar Field Wormhole}
\author{Tonatiuh~Matos$^{1,2}$ and Dar\'{\i}o~N\'{u}\~{n}ez$^3$}
\affiliation{$^{1}$CIAR Cosmology and Gravity Program, Department
of Physics and Astronomy, University of British Columbia,
Vancouver, British Columbia, Canada, V6T 1Z1\\ $^{2}$Departamento
de F{\'\i}sica, Centro de Investigaci\'on y de Estudios Avanzados
del IPN, A.P. 14-740, 07000 M\'exico D.F., M\'exico.}
\email{tmatos@fis.cinvestav.mx}
\affiliation{$^{3}$Instituto de Ciencias Nucleares, Universidad
Nacional Aut\'onoma de M\'exico, Apdo. 70-543, CU, 04510 M\'exico,
D.F., M\'exico.} \email{nunez@nucleares.unam.mx}

\begin{abstract}
We derive an exact solution of the Einstein's equations with a
scalar field stress-energy tensor with opposite-sign, and show
that such a solution describes the inner region of a rotating
wormhole. We also show that the non-rotating case of such a
solution represents a static, asymptotically flat wormhole
solution. We match the radial part of the rotating solution to the
static one at both mouths, thus obtaining an analytic description
for the asymptotic radial region of space-time.
We explore some of the features of these solutions.
\end{abstract}

\date{\today}



\maketitle

\section{Introduction}
The wormhole (WH) solutions of the Einstein equations started with
Einstein himself, since he was interested in giving a field
representation of particles \cite{ER}. The idea was further
developed by Ellis, \cite{Ellis} and others, where instead of particles,
they try to model them as "bridges" between two regions of the
space-time. The idea of considering such solutions as actual
connections between two separated regions of the Universe has
attracted a lot of attention since the seminal work of Morris and Thorne
\cite{MT}. These solutions have evolved from a
sort of science-fiction scenario, to a solid scientific topic, even to
the point of considering the actual possibility of their existence.

Indeed, Carl Sagan, worried about the impossibility of star travel
imposed by the fundamental laws of Physics, the most serious
problems being the huge distances, coupled with the finiteness of the
speed of light and Lorentz time contraction, asked physicist for
help. It turned out that the solution proposed by
Ellis \cite{Ellis} actually could be interpreted as the
identification, or union, of two different regions, no matter how
far apart they were or even if those two regions were in the same
space-time. You could still identify two regions. He obtained a
solution which allows one to go from one region to another by means
of this identification.

In any case, such solutions need an "exotic" type of matter which violates the energy conditions
which we are usually obeyed. See \cite{Viser1}
for a detailed review of this subject. The solutions do exist
but they need to be generated by matter which apparently does not
exist. Actually one needs something very peculiar to warp
space-time or to make holes in it. This feature was a serious
drawback to the possibility of their actual existence in nature, so WH's
remained in the realm of fiction.

However, as often happens with these "impossible"
conjectures, more and more evidence has appeared pointing toward the
presence in our Universe of unknown types of matter and energy
which do not necessarily obey the usual energy conditions. Indeed,
it is known that the
Universe contains $73 \%$ of dark energy. This
new type of matter forms the overwhelming majority of matter in
the Universe and seems to be found everywhere \cite{EO}. Some works have
also discussed the plausibility of energy-condition
violations at the quantum level (see for example \cite{Rom}). There
is now agreement in the scientific community that matter
which violates some of the energy conditions may very possibly
exist. Thus, the idea that the WH's can be rejected because of the
type of matter that is needed for them to exist is not as tenable as it once was.

Another mayor problem faced by WH solutions is their
stability. By construction, WH solutions are traversable. That
is, a test particle can go from one side of the throat to the
other in a finite time as measured by an observer on the test
particle and by another observer far away from it, and without feeling large
tidal forces. The stability problem of the "bridges" has been
studied since the $60's$ by Penrose \cite{Pen} in connection to
the stability of Cauchy horizons. However, the stability of
the throat of a WH was just recently studied numerically by
Shinkay and Hayward \cite{shin}. They show that the WH proposed by
Thorne \cite{MT}, when perturbed by a scalar field with a
stress-energy tensor with the usual sign, the WH may possibly collapse
to a black hole and the throat closes. In the same way,
when the perturbation is due to a scalar field of the same type as
that generating the WH, the throat grows exponentially, showing
that the solution is highly unstable.

Intuitively it is clear that a rotating solution would have a higher
possibility of being stable, and so would static
spherically symmetric solutions more general than the one proposed by Thorne.
Some studies of rotating WH solutions exist \cite{rot},
but no one has found an exact solution to the Einstein
equations describing such a WH. In the present work we do so.

\section{Field equations}
In a nutshell, the idea  was to use solution generation
techniques developed in the late $80's$ for the Einstein's
equations (see \cite{Kram}), where it was possible, in the
chiral formulation \cite{Matos}, to derive the Kerr solution
starting from Schwarzschild, so we decided to use the
same techniques applied to the WH proposed by Ellis and Thorne.
The details of the derivation will be presented elsewhere, but the
final result was the following ansatz for the line element:
\bea ds^2&=&-f\left(c\,dt+a\,\cos\theta\,d\varphi\right)^2
 \nonumber \\
&+&\frac{1}{f}\,\left[dl^2+\left(l^2-2\,l\,l_1+{l_0}^2\right)\,
\left(d\theta^2+\sin^2\theta\,d\varphi^2\right)\right],
\label{eq:elem} \eea
where $f=f(l)$ is an unknown function to be determined by the
field equations, $l_0, l_1$ are constant parameters with units of
distance such that $l_0^2 > l_1^2$, and $l_0 \neq 0$, $c$ is the
speed of light and $a$ is the rotational parameter (angular
momentum per unit mass). In these coordinates the distance $l$
covers the complete manifold, going from minus to plus infinity.
Notice that, modulo $f$, there is already the throat or bridge
feature of the WH's in such a line element, as the coefficient of
the angular variables is never zero.

The constant parameter $a$ is indeed a rotation parameter, as we show
for the exact solution presented bellow.

The process of formation of a WH is still an open question. We
suppose that some scalar field fluctuation collapses in such a way
that it forms a rotating scalar field configuration. This
configuration has three regions; the interior, where the
rotation is non-zero and two exterior regions, one on each side
of the throat, where the rotation stops. The inner boundaries of
this configuration are defined where the rotation vanishes. The
interior field is the source of the WH and we conjecture that its
rotation will keep the throat from being unstable. In
what follows we construct such an exact solution.

We look for a solution to the Einstein equations with an
stress-energy tensor describing an opposite sign massless scalar
field, $\phi$ (see \cite{phantom}), so that the field equations
take the form
\be R_{\mu\nu}=-\frac{8\,\pi\,G}{c^4}\,\phi_{,\mu}\,\phi_{,\nu},
\label{eq:Einstein} \ee
with $R_{\mu\nu}$ the Ricci tensor. It is quite remarkable that
for such an ansatz described by Eq.~(\ref{eq:elem}), the field
equations take a very simple form. Demanding stationarity for the
scalar field, it turns out that it can only depend on the distance
coordinate $l$. We are left with only two Einstein equations. For
simplicity we work with the components $^t_t$ and the sum of
$^t_t$ and $^r_r$ of equation (\ref{eq:Einstein}), we obtain:
\bea
\left(\left(l^2-2\,l_1\,l+{l_0}^2\right)\,\frac{f^\prime}{f}\right)^\prime
+
\frac{{a}^2\,f^2}{l^2-2\,l_1\,l+{l_0}^2}&=&0, \label{eq:ecua0}\\
\left(\frac{f^\prime}{f}\right)^2 +
\frac{4\left({l_0}^2-l_1^2\right)+{a}^2\,f^2}{\left(l^2-2\,l_1\,l+{l_0}^2\right)
^2} -\frac{16\,\pi\,G}{c^4}\,{\phi^\prime}^2&=&0,\label{eq:ecua}
\eea
where prime stands for the derivative with respect to $l$. Recall
that the Klein Gordon equation is a consequence of the Einstein's
ones, so we only need to solve (\ref{eq:ecua0}),(\ref{eq:ecua}).
As described above, we separate the solutions in the inside and
the outside ones.

\section{The Static Solution}

\begin{figure}[htb]
\includegraphics[width=3cm]{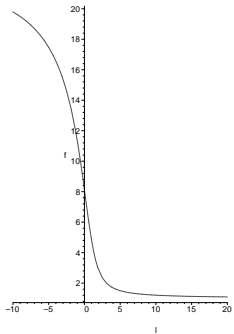}
\caption{\label{fig:f} Plot of the function $f$ at the positive
and negative regions near the throat. Notice that $f$ is smooth
and positive everywhere. The values of the parameters are $l_0 =
2, l_1 = 1, \phi_0 = 1$.}
\end{figure}

The static case for such an ansatz, Eq.(\ref{eq:elem}) with $a=0$,
can be solved in general, giving the following solution:
\bea f_{\rm st}&=&e^{-\phi_0\,\left(\lambda-\frac{\pi}{2}\right)},
\label{solucion_estatica1} \\
\sqrt{\frac{8\pi G}{c^4}}\phi_{\rm
st}&=&\sqrt{2+\frac{{\phi_0}^2}{2}}\,
\left(\lambda-\frac{\pi}{2}\right),\label{solucion_estatica2}\\
\lambda&=&\arctan\left(\frac{l-l_1}{\sqrt{{l_0}^2-{l_1}^2}}\right),
\label{solucion_estatica} \eea
where $\phi_0$ is an unitless integration constant. The scalar
field is given as multiples of the Plank's mass. Notice that this
is by itself an exact solution for the complete manifold. It is
spherically symmetric, and recalling that $\arctan(x)$ goes to
$\frac{\pi}{2}$ for $x$ tending to infinity, we have chosen the
other integration constants in such a way that the scalar field
vanishes for large positive values of $l$, and $f$ tends to one
for those values. On the other side of the throat, for large
negative $l$, the space-time is also flat but is described in
different coordinates. The time and distance intervals on both
sides and far from the throat are given by:
\bea {ds^2}_{\rm far}&=&-dt_{+}^2+dl_{+}^2 +l_{+}^2d\Omega^2\nonumber\\
&=&-e^{{\phi_0\,\pi}}dt_{-}^2+
\frac{1}{e^{{\phi_0\,\pi}}}\left(dl_{-}^2+l_{-}^2d\Omega^2\right).
\eea
Thus, intervals on both sides of the throat are related by
\bea
\Delta\,t_+&=&e^{\frac{\phi_0\,\pi}{2}}\,\Delta\,t_-, \\
\Delta\,l_+&=&e^{-\frac{\phi_0\,\pi}{2}}\,\Delta\,l_-, \eea
which is an interesting relation. For a large positive value of
$\phi_0$, a short interval in time for the observer at the
positive side is seen as a large time interval at the negative
side, and short distances seen in the negative side are measured
as large distances in the positive side. Also, the value of the
scalar field at the negative side is constant but not zero, as the
zero value was chosen at the other side. In figure Fig.\ref{fig:f}
we plot the gravitational function $f$ for positive and negative
values of the distance coordinate near the throat.

Expanding the gravitational function $f_{\rm st}$ and the scalar
field $\phi_{\rm st}$ for large values of the distance coordinate
$l$, we have, on the positive side:
\bea f_{\rm st}&=&1+\frac{\phi_0\,\sqrt{l_0^2-l_1^2}}{l} +
O\left(\frac{1}{l^2}\right)\\
\sqrt{\frac{8\pi G}{c^4}}\phi_{\rm
st}&=&-\sqrt{2+\frac{\phi_0^2}{2}}\,\frac{\sqrt{l_0^2-l_1^2}}{l} +
O\left(\frac{1}{l^2}\right) \label{estatica_expansion} \eea
\noindent from where we can read the mass parameter which takes
the form $M=-\frac{\phi_0\,\sqrt{l_0^2-l_1^2}}{2}$ and the scalar
field charge $\sqrt{\frac{8\pi
G}{c^4}}q_{\phi}=-\sqrt{2+\frac{\phi_0^2}{2}}\,\sqrt{l_0^2-l_1^2}$.
Notice how the mass parameter depends on the value of $\phi_0$
and is positive for negative $\phi_0$, the scalar field charge is
invariant with respect to the sign of $\phi_0$

\begin{figure}[htb]
\includegraphics[width=4cm]{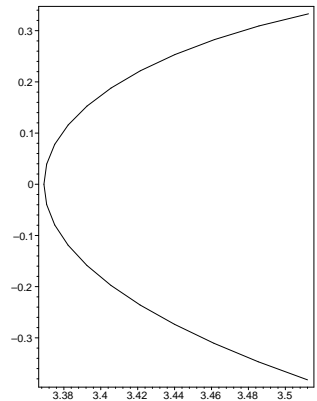}
\caption{\label{fig:g_esf} Shape of the throat for the static WH.
The values of the parameters are $l_0 = 3.54, l_1 = 1, \phi_0 = 0.01$.
For these values, the minimal length of the throat is $\rho_{\rm
min}=3.3692$, for $l_{\rm min}=0.9830$, and we plot for $l_{\rm
min}-1$, to $l_{\rm min}+1$.}
\end{figure}

Now, with respect to the throat, following Morris and Thorne
\cite{MT}, we freeze the time and select the plane angle
$\theta=\frac{\pi}{2}$ in that slice obtaining the ${ds_2}^2=
\frac{1}{f}\,\left(dl^2+\left(l^2-2\,l\,l_1+{l_0}^2\right)\,d\varphi^2\right)$.
Making the embedding in the cylindrical space
$dD^2=d\rho^2+dz^2+\rho^2\,d\varphi^2$, the shape of the throat is
given by $z=z(\rho)$, thus
${dD_2}^2=\left(1+{z_{,\rho}}^2\right)\,d\rho^2+\rho^2\,d\varphi^2$.
Comparing the two distant elements we see that
$\rho^2=\frac{l^2-2\,l\,l_1+{l_0}^2}{f}$, and the shape of the
throat is given then by $z(\rho)=\int_{\rho_{\rm
min}}^\rho\,\sqrt{\frac{(l_{,\rho})^2}{f(\rho)}-1}\,\,d\rho$. In
our case, it turns out to be more convenient to do this procedure
in a parametric form, in terms of the distance coordinate $l$, so
the value for the throat is given by the pair
$\left[\rho(l_f),z(l_f)=\int_{l_{\rm
min}}^{l_f}\,\sqrt{\frac{1}{f}-(\rho_{,l})^2}\,\,dl\right]$. Thus,
given an $l_f$, we obtain the value of the corresponding pair of
coordinates. The integral was solved numerically, the resulting
shape of the throat is given in Fig.\ref{fig:g_esf}. Notice the
distinctive shape of the throat.

Solution
(\ref{solucion_estatica1},\ref{solucion_estatica2},\ref{solucion_estatica})
includes as particular solution the Ellis-Morris-Thorne one
described in terms of the proper distance. This can be seen by
taking $\phi_0=0, l_1=0$, which reduces the static solution to
$f=1$ and the scalar field to $\sqrt{\frac{8\pi
G}{c^4}}\phi=\sqrt{2}\,\left(\lambda-\frac{\pi}{2}\right)$, with
$\lambda=\arctan\left(\frac{l}{l_0}\right)$, which is just the
well known static spherical symmetric solution of
Ellis-Morris-Thorne,\cite{MT,Ellis}.

In this way, we see that the WH static solution is, by itself, an
interesting solution. A deeper study of it, including the geodesic
motion of particles, the description of the physical parameters
acting on it, as well as a stability analysis, are subjects beyond
the aim of the present work, which is to obtain exact solutions to
the Einstein-scalar field equations. Thus, we continue to present
the case with the rotation parameter $a\neq0$ for which, as we will
show, the field equations can also be solved, describing a
rotating wormhole with a deficit angle in the asymptotic regions,
and can also be thought to model the region near the throat of a
rotating WH. We show that it can be smoothly matched with the
asymptotically flat external regions, in particular we match the
radial part of it to the static solution presented above.
\begin{figure}[htb]
\includegraphics[width=9cm]{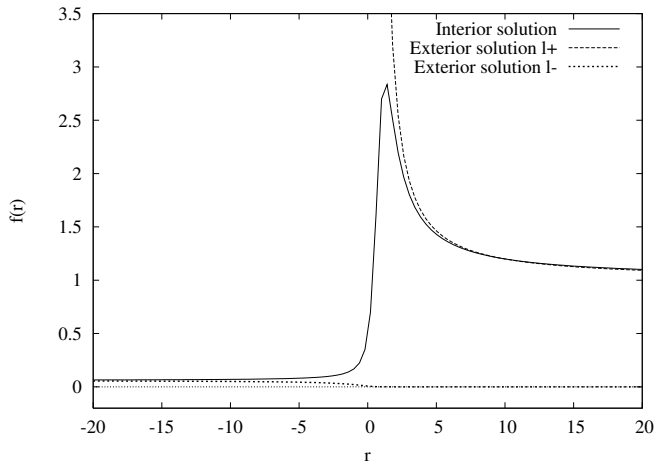}\\
\caption{\label{fig:pegado} Plot of the function $f$ for the
exterior and interior solutions, observe that for these values the
matching of the solutions on the rhs is smooth, while the matching
is always smooth on the lhs. The values of the parameters are $l_1
= 0.5, l_{+} = 10, \phi_0 = 2, l_0 = 1, a = 0.6$ and we plot for
radius from -20 to 20.}
\end{figure}

\section{The Rotating Solution}
For $a\neq0$ we find the exact solution:
\bea f&=&2\,{\frac {{\phi_0}\,\sqrt {D\, \left(
{l_0}^{2}-{l_1}^{2}
 \right) }\,{e^{{ \phi_0}\, \left( \lambda-\frac{\pi}{2} \right) }}}{{a
}^{2}+{D}\,{e^{2\,{\phi0}\, \left( \lambda-\frac{\pi}{2}
 \right) }}}}
, \label{eq:f}\\
\sqrt{\frac{8\pi
G}{c^4}}\phi&=&\sqrt{2+\frac{{\phi_0}^2}{2}}\,
\left(\lambda-\frac{\pi}{2}\right)
, \eea
where the function $\lambda$ is again given by
(\ref{solucion_estatica}), $a$ is the angular momentum parameter
and $D$ is an integration constant. It is remarkable that the
scalar field has exactly the same expression as in the
non-rotating case. Here $\phi_0>0$ in order to preserve the
signature of the metric. It can be seen that there is a throat,
 for large values of the
distance parameter, $l$, the gravitational functions goes to a
constant on both sides, as well as the scalar field, showing that
we do have a rotating wormhole solution to the field equations,
with a deficit angle. Furthermore, the norm of the Killing-vector
$\xi=\frac{\partial}{\partial t}$ is always negative $|\xi|=-f$,
thus the solution does not have an ergosphere. As mentioned
before, this is the first solution of this type, as long as the
other ones known in the literature, \cite{rot}, only describe the
possible geometry of a rotating wormhole, without solving the
complete field equations with a specific source of matter.

We can see that indeed the constant $a$ is a rotation parameter by
means of the Ernst potential $\mathcal{E}$ (see for example
\cite{exactsolutions}). Due to the symmetries of the space-time,
it is easy to see that the Ernst potential is
$\mathcal{E}=f+i\epsilon$, where $\epsilon$ is the rotations
potential. For solution (\ref{eq:f}) the invariant quantity
$\epsilon$, is given by
\be \epsilon=-a\frac{\phi_0\sqrt{D(l_0^2-l_1^2)}}{a^2+De^{2\phi_0(\lambda-\frac{\pi}{2})}}, \ee
showing in this way that the parameter $a$ is a parameter of the
space-time as claimed.

Even though this is a solution for the complete space-time, we can
see it as describing the internal parts of a rotating wormhole,
from the throat to some distance on each mouth, and match it with
an external, asymptotically flat solution, such as Kerr or the
static wormhole presented above. In what follows we present this
last matching for the radial part of the solution, showing that it
can be a smooth one. Then, the matched function for the rotating
WH solution reads
\be f=\left\{
\begin{array}{cc}
\exp (\lambda -\frac{\pi }{2}) & \text{if }l>l_{+} \\
&  \\
2\,{\frac{{\phi _{0}}\,\sqrt{D\,\left(
{l_{0}}^{2}-{l_{1}}^{2}\right) }\,{e^{{ \phi _{0}}\,\left( \lambda
-\frac{\pi }{2}\right) }}}{{a}^{2}+{D}\,{e^{2\,{ \phi_0}\,\left(
\lambda -\frac{\pi }{2}\right) }}}} & \text{if }l_{-}\leq l \leq
l_{+}
\\
&  \\
\phi _{1}\exp (\lambda +\frac{\pi }{2}) & \text{if }l<l_{-}
\end{array}
\right. \label{eq:solucion_rot} \ee
where $l_{-}$ and $l_{+}$ respectively are the matching points on
the left and right hand side (see Fig.\ref{fig:pegado}). It can be
seen that the interior solution matches with the rhs exterior one
provided that the parameter $D$ becomes:
\bea {D}&=&2\,{\phi_0}\,\sqrt { \left( {l_0}^{2}- {l_1}^{2}
\right) \left( \left( {l_0}^{2}-{l_1}^{2} \right)
{{\phi_0}}^{2}-{\frac {{a}^{2}}{{E}^{2}}} \right) }\nonumber \\
&+&2\, \left( {l_0}^{2}-{l_1}^{2} \right) {{\phi_0}}^{2}-{ \frac
{{a}^{2}}{{E}^{2}}} \eea
\noindent where the constant $E$ is determined by the radio
$l_{+}$ where the two solutions match, it is given by:
\be E=\exp\left[{{\phi_0}\left(\arctan\left({\frac
{{l_{+}-l_1}}{\sqrt{l_0^2-l_1^2}}}
 \right) - \frac{\pi}{2} \right)}\right].
\ee

In order to have a real solution everywhere we impose the
constraint that $4\,M^2=\left( {l_0}^{2}-{l_1}^{2} \right)
{\phi_0}^{2}>\frac {{a}^{2}}{{E}^{2}}$. On the other side the
matching of the interior solution is smooth with the lhs exterior
one, if the constant $\phi_1$ is chosen such that
\be {\phi_1}=2\,{\frac {{\phi_0}\,\sqrt
{D({l_0}^{2}-{l_1}^{2})}\,e^{2\,\phi_0\,\lambda_-
}}{{a}^{2}+{D}{e^{2\,{\phi_0}\,\left(\lambda_- - \frac{\pi}{2}\right) }}}}, \ee
where $\lambda_-$ is given by Eq.~(\ref{solucion_estatica}),
evaluated at $l=l_-$. In Fig.\ref{fig:pegado} we see the plot of
$f$. Observe that the matching on the rhs could be smooth if
$l_{+}$ is sufficiently large and the rotation parameter $a$ is
sufficiently small. On the contrary, for small $l_{+}$ or/and big
rotation parameter the matching on the rhs is continous but not
necessarily smooth. On the lhs and for the scalar field the
matching is always smooth. Following the same procedure described
above to obtain the shape of the throat, we obtain a similar
figure like Fig.\ref{fig:g_esf} as for the non-rotating case.

\section{Physical Aspects of the solution}

Let us write the solution in terms of the mass parameter $M$. We
start with the constant $D$, it reads
\bea D&=&M^{2}\ \left(4\,\sqrt { 4-{\frac {{J}^{2}}{{E}^{2}}}
}+8-{
\frac {{J}^{2}}{{E}^{2}}}\right)\nonumber\\
&=&M^{2}\ d^2 \eea
where $J=a/M$. Observe that $d$ does not depend on the mass
parameter $-M=mG/c^2$, where $m$ is the total mass of the scalar
field star. Furthermore, if we want that the wormhole is
transversable, we expect a gravitational field similar to the
Earth's one, this means $m\sim$ Earth's mass. For this value the
mass parameter is $-M\sim 0.01\ $ meters. But we want that a
spaceship can go trough the throat of the wormhole, we can suppose
that $l_{0}\sim 10$ meters, then, $l_{1}\sim 10$ meters as well,
depending on the value of $\phi_{0}$. Thus, the interior solution
reads
\bea f_{int}&=& 4\,{\frac{-M\,\sqrt{D }\,{e^{{ \phi _{0}}\,\left(
\lambda -\frac{\pi }{2}\right) }}}{{a}^{2}+{D}\,{e^{2\,{
\phi_0}\,\left( \lambda -\frac{\pi }{2}\right) }}}}\nonumber\\
&=& 4\,{\frac{d \,{e^{{ \phi _{0}}\,\left( \lambda -\frac{\pi
}{2}\right) }}}{{J}^{2}+d^2\,{e^{2\,{ \phi_0}\,\left( \lambda
-\frac{\pi }{2}\right) }}}} \eea
where the function $\lambda$ is given by
\be \lambda=\arctan\left(-\phi_{0}\frac{l}{2M}\right) \ee
Observe that $l_{Sch}=-2M$ is the Schwarzschild radius. The scalar
field star which provokes the wormhole could be thousands of
meters long. Compare with the mass parameter $M$ of orders of
millimeters, the matching could be considered as if it were at
infinity. In that case the parameter $E=1$. Thus the solution has
only 3 parameters, $M,\ J$ and $\phi_{0}$. In this unites $|J|\leq
2$. Then the parameter $d$ is bounded to $2\leq d\leq4$.
Nevertheless, the matching condition on the left hand side (in
$l_{-}$) does not depend on the mass parameter $M$, it is given by
\be \phi_{1}=
4\,{\frac {d\,{e^{2\,\phi_0\,\lambda_-}}}{{J}^{2}+d^2{e^{2\,\phi_0\,\left(\lambda_- - \frac{\pi}{2}\right)}}}}. \ee
where $\lambda_-=\lambda(l_-)$. We can take $l_-$ such that
$\lambda_-\simeq-\pi/2$. Thus, on the left hand side the metric at
minus infinity can be written as
\be ds^2_{\rm
far}=-\phi_{1}dt_{+}^2+\frac{1}{\phi_{1}}\left(dl_{+}^2
+l_{+}^2d\Omega^2\right) =-dt_{-}^2+dl_{-}^2+l_{-}^2d\Omega^2, \ee
where now $\Delta t_{-}=\sqrt{\phi_{1}}\Delta t_{+}$ and $\Delta
l_{-}=\frac{1}{\sqrt{\phi_{1}}}\Delta l_{+}$.
The constant $\phi_{1}$ can be very big. Amazingly the major values
for $\phi_{1}$ are obtained for small rotation, $J<<1$. It can be seen that
$\phi_{1}$ has a maximum value when $\phi_{0}$ is
\be \phi_{0 max}=-\frac{1}{2\pi}\ln  \left( {\frac
{{J}^{2}{E}^{2}}{4\,\sqrt { 4\,E^{2}-J^{2}}\
E+8\,{E}^{2}-{J}^{2}}} \right) \ee
Thus, an observer on the left hand side space (on $l_{-}$) will
feel that the time goes fast for small changes in $\Delta t_{+}$
and will measure small changes of space for big changes on
$\Delta l_{+}$. It is left to find a phantom field star
with small rotation and scalar charge given by
\be \sqrt{\frac{8\pi
G}{c^4}}q_{\phi}={2M}\sqrt{\frac{2}{\phi_{0}^{2}}+\frac{1}{2}}
\ee
with a scalar charge given by $\phi_{0}=\phi_{0max}$. For example,
if the phantom scalar star has a rotation like $J=10^{-10}$, then
$\phi_{1}\sim 1.4\times10^{5}$ when the scalar charge is
$q_{\phi}\sim 0.3M\ m_{Plank}$. For a star with an Earth's mass,
this charge is equivalent to $q_{\phi}\sim 0.003\ m_{Plank}$ per
meter.

\section{Stability of the Solutions}

Thus far we have shown that we have indeed solved the field
equations and obtained a rotating wormhole, with a static one as a
particular case, and presented some physical properties of the
solutions. This is the main purpose of the present work. As long
as the stability issue is of the most exciting ones, we finish the
work with an idea that gives us ground to conjecture that the
rotating solution is stable indeed. Of course, the complete
analysis, needs a numerical evolution of the perturbed solution,
or an analysis of the radiative modes in the Teukolsky equation.
Here we present some preliminary results concerning the stability
of the solutions. Let $h_{\mu\nu}=h_{\mu\nu}(l,t)<<1$ be a radial
perturbation of metric (\ref{eq:elem}), that is
\bea ds^2&=&-(f+h_{44})c^2\,dt^2-2\ a\,\cos\theta\ dt\ d\varphi
 \nonumber \\
&+&\left(\frac{1}{f}+h_{11}\right)\,
dl^2+\frac{1}{f}\left(l^2+{l_0}^2\right)\,
d\theta^2\nonumber\\
&+&\left(\frac{1}{f}\left(l^2+{l_0}^2\right)\,\sin^2\theta-f\ a\
\cos\theta\right)\,d\varphi^2, \label{eq:elem_pert} \eea
(In this section we make the distance parameter $l_{1}=0$, for
simplicity). Now, let us consider a Gaussian ring perturbation in
the radial direction centered in $l=L_{0}$, for $\theta=\pi/2$, in
such a way that $h_{11}=f_{0}T(t)\exp(-(l-L_{0})^2/L_{1})$. Even
for this simple perturbation, the field equations are a
non-trivial set of differential equations. However, in this case
the behavior of the field equations can be reduced to an evolution
equation only for $h_{11}$ of the form
\be T_{,tt}+\omega^2\ T+\text{terms} \ee
\begin{figure}[htb]
\includegraphics[width=9cm]{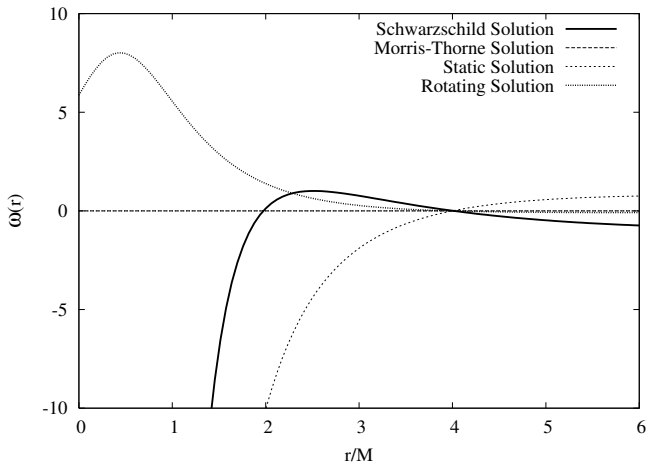}
\caption{\label{fig:estabil} Plot of the function $\omega$ for the
Schwarzschild, the Morris-Thorne, the static and the rotating
solutions. The Schwarzschild solution has unstable modes only
inside the horizon $l=-2M$, but it has stable modes outside, as
the rotating solution has. We have set the ring perturbation as
$r_{0}=4\ r_{1}=0.1$ and $f_{0}=1\times10^{-4}$. }
\end{figure}
where $\omega$ is a function of the radial coordinate $l$ only.

Not pretending to give a demonstration of stability, we can expect
that this evolution equation will be oscillating if $\omega^2>0$,
and monotonic if $\omega^2<0$. In Fig.\ref{fig:estabil} we see the
results using this criterion for the Schwarzschild, the
Morris-Thorne, the static (\ref{solucion_estatica1}) and the
rotating (\ref{eq:solucion_rot}) solutions. We can interpret these
results as follows. We have added a perturbation to the solution in
such a way that the Gaussian perturbation simulates the presence
of an object in $l=L_{0}$ and remains there. From the curve
obtained from the Schwarzschild solution we see that the
perturbation provokes monotonic modes for the region $l>L_{0}$,
but oscillating modes for the region $l<L_{0}$. Inside the horizon
the modes are again monotonic, this does not affect the stability
of the object. This might imply that the original object is
stable, the perturbed metric oscillates around the original metric
for the region inside of the center of the perturbation on
$l=L_{0}$. This situation does not happen with the Morris-Thorne
solution. There, $\omega=0$ implying that all the modes are
monotone everywhere. This can be interpreted as saying that this
solution is unstable for these kind of perturbations. In the case
of the static solution (\ref{solucion_estatica1}) the modes are
monotone for the region $l<L_{0}$, but it is oscillating for the
regions $l>L_{0}$. This might imply that the static solution is
unstable as well, because the monotonic behavior of the solution
close to the center will destroy the object. For the case of the
rotating solution (\ref{eq:solucion_rot}) the situation is similar
as for the Schwarzschild one. The modes in the region $l<L_{0}$
are oscillating, the perturbed solution on this regions only
oscillates around the original solution. We see that, according to
this criterion, the Schwarzschild solution is stable under this
Gaussian perturbation, but the Morris-Thorne and the static are
not, and that the rotating solutions is stable. Nevertheless,
because this is not a demonstration in general, we leave this
result as a conjecture.

If this would be so, then the possible existence of negative
kinetic energy scalar fields and its stability could be the door
to a new way of travel form through the Universe.

\section{Conclusions}

We have found a new solution which represents the space-time of a
rotating wormhole. This is the first exact solution of the
Einstein equations with a scalar field, with negative kinetic term
(phantom field), with these features. The solution is matched to
an static one which is by self an wormhole solutions with a
phantom field source. The solutions present the expected features,
they connect to regions of the space-time, where time and space
behave in different way in both regions. We found that the
different behavior depends on the parameters of the solution, $J,\
M$ and $q_{\phi}$. Slow rotating wormhole stars contain more
possibilities to connect a human traveller with more distant
regions. We present some calculations of the perturbations which
could indicate that the rotating solution is stable under certain
perturbations. We consider that counting with an exact solution to
the Einstein scalar field equations for a rotating wormhole will
certainly contribute to have a better understanding on the
physical processes occurring in those regions, suggesting what
kind of observations should be made in order to probe the
existence or not of wormholes in the Universe.


\section{Acknowledgments}
The authors acknowledges Miguel Alcubierre, Olivier Sarbach,
Roberto Sussman, Juan Carlos Degollado, Mike Ryan, Abril Suarez
and Israel Villagomez for fruitful comments on our work. TM wants
to thank Matt Choptuik for his kind hospitality at the UBC. DN is
grateful to Vania Jimenez-Lobato for useful discussions to start
the elaboration of the present work. This work was partially
supported by DGAPA-UNAM grant IN-122002, and CONACyT M\'exico,
under grants 32138-E and 42748.

\end{document}